\documentclass{iaus}
\usepackage{graphicx}

\title[LOFT: Large Observatory For X-ray Timing] 
{LOFT: Large Observatory For X-ray Timing}

\author[R. P. Mignani, S. Zane, et al.]   
{R. P. Mignani$^{1,2}$,
S. Zane$^{1}$,
D. Walton$^{1}$,
T.  Kennedy$^{1}$,
B.  Winter$^{1}$,
P.  Smith$^{1}$,
R. Cole$^{1}$,
D. Kataria$^{1}$,
A. Smith$^{1}$,
on behalf of the {\em LOFT} team
}

\affiliation{$^1$ Mullard Space Science Laboratory, University College London, \\ Holmbury St. Mary, Dorking, Surrey, RH5 6NT, UK \\email: {\tt rm2@mssl.ucl.ac.uk} \\ [\affilskip]$^2$ Kepler Institute of Astronomy, \\ University of Zielona G\'ora, Lubuska 2, 65-265, Zielona G\'ora, Poland }

\pubyear{2011}
\volume{285}  
\pagerange{119--126}
\jname{New Horizons in Time Domain Astronomy}
\editors{E. Griffin,  R. Hanisch, \& R. Seaman eds.}
\begin{document}

\maketitle

\begin{abstract}
High-time-resolution X-ray observations of compact objects provide direct access to strong field gravity, black hole masses and spins, and the equation of state of ultra-dense matter. {\em LOFT}, the large observatory for X-ray timing, is specifically designed to study the very rapid X-ray flux and spectral variability that directly probe the motion of matter down to distances very
close to black holes and neutron stars. A 10 m$^2$-class instrument in combination with good spectral resolution ($<$260 eV @ 6 keV) is required to exploit the relevant diagnostics and holds the potential to revolutionise the study of collapsed objects in our Galaxy and of the brightest supermassive black holes in active galactic nuclei. {\em LOFT} will carry two main instruments: a Large Area Detector (LAD), to be built at MSSL/UCL with the collaboration of the Leicester Space Research Centre for the collimator) and a Wide Field Monitor (WFM). The ground-breaking characteristic of the LAD (that will work in the
energy range 2--30 keV) is a mass per unit surface in the range of $\sim$10 kg/m$^2$, enabling an effective area of $\sim$10 m$^2$ (@10 keV) at a reasonable weight and improving by a factor of $\sim$ 20 over all predecessors. This will allow timing measurements of unprecedented sensitivity, allowing  the capability to measure the mass and radius of neutron stars with $\sim$5\% accuracy, or to reveal blobs orbiting close to the marginally stable orbit in active galactic nuclei. In this contribution we summarise the characteristics of the {\em LOFT} instruments and give an overview of the expectations for its capabilities.

\end{abstract}

\firstsection 
              
\section{A New X-ray Mission}
             
{\em LOFT}\footnote{http://www.isdc.unige.ch/loft/index.php/the-loft-mission} is one of four M3 missions that have been selected by ESA for an Assessment Phase and be  considered for a possible launch in 2020-2022. The {\em LOFT} Consortium includes several institutes across the UK, Europe, Israel, Turkey, Canada, the US, and Brazil. In addition to MSSL, the UK participation include the Space Research Centre (SRC) in Leicester, the University of Southampton,  the University of Durham, the University of Manchester, and the University of Cambridge. The UK participation is sponsored by the  UK Space Agency. MSSL/UCL will lead the {\em LOFT} Large Area Detector (LAD) instrument within the consortium (S. Zane, D. Walton) as well as will have a major role in the hardware/software development and system engineering (thermal, mechanical, electronics and software). This effort will be supported by Leicester SRC (G. Fraser) who lead the development of the collimators. Southampton, Durham. Manchester, and Cambridge are among the other UK institutions currently involved in the {\em LOFT} Science consortium.

{\em LOFT} (Feroci et al.\ 2011a,b) is a 10 m$^2$-class telescope, specifically designed to study the very rapid X-ray flux and spectral variability that directly probe the motion of matter down to distances very close to black holes and neutron stars. 
High-time-resolution X-ray observations of compact objects provide direct access to strong-field gravity, black hole masses and spins, and the equation of state of ultra-dense matter. They provide unique opportunities to reveal for the first time a variety of general relativistic effects, and to measure fundamental parameters of collapsed objects. They gain unprecedented information on strongly curved space times and matter at supra-nuclear densities and in supercritical magnetic fields. In turn, this bears directly to answer several fundamental questions of both the ESA's Cosmic Vision Theme  ÒMatter under extreme conditionsÓ and the STFC road map ÒWhat are the law of the physics under extreme conditions?Ó. 
A 10 m$^2$-class telescope as {\em LOFT},  in combination with good spectral resolution ($<$260 eV @ 6 keV) is required to exploit the relevant diagnostics and holds the potential to revolutionise the study of collapsed objects in our galaxy and of the brightest supermassive black holes in active galactic nuclei (AGNs). The timescales/phenomena that {\em LOFT} will investigate range from sub-millisecond, quasi-periodic oscillations to year-long transient outbursts, and the relevant objects include many that flare up and change state unpredictably. Thus, relatively long observations, flexible scheduling and continuous monitoring of the X-ray sky are essential elements for success.

\begin{figure}
\begin{center}
 \includegraphics[width=2.0in]{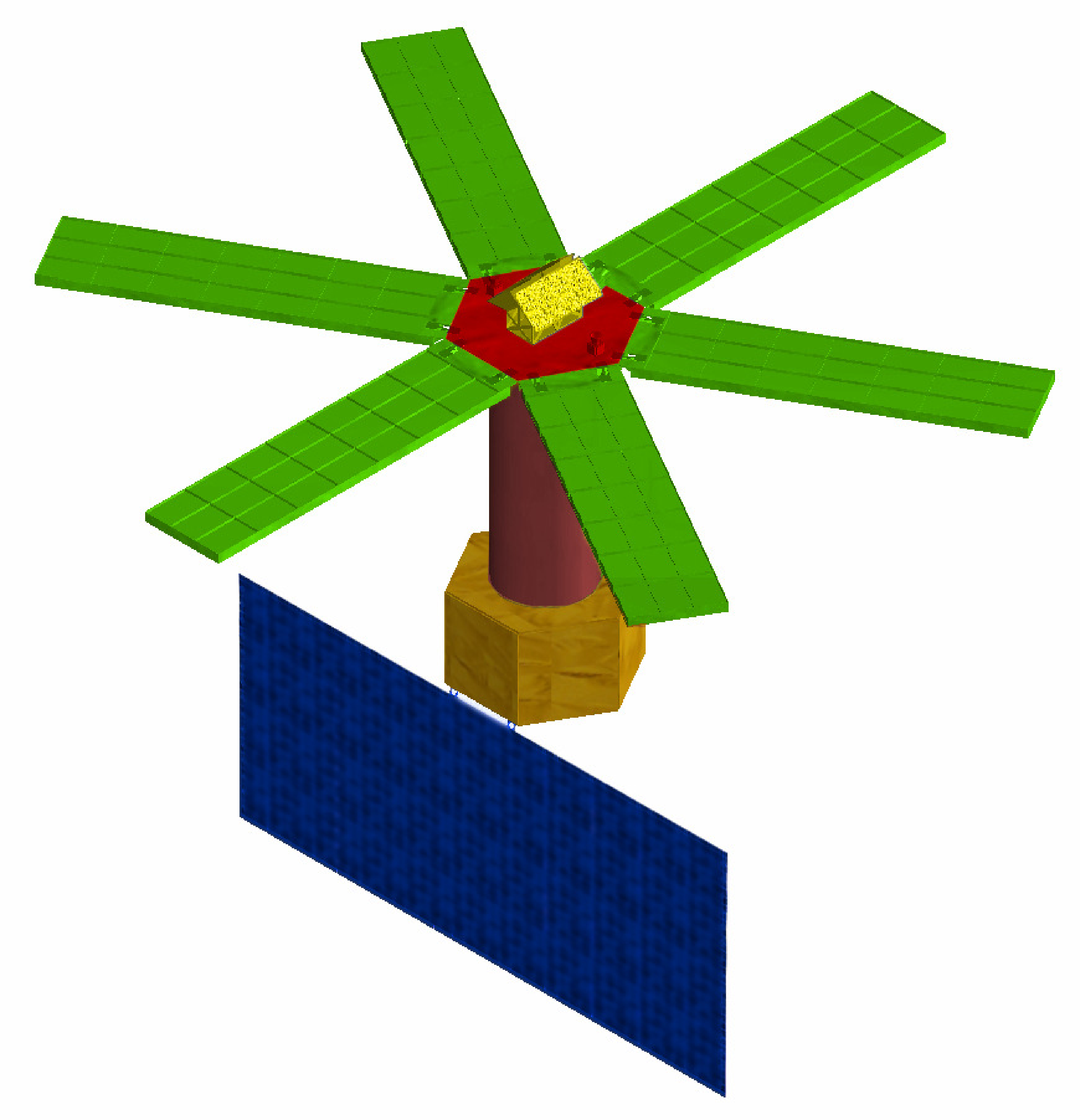} 
  \includegraphics[width=2.8in]{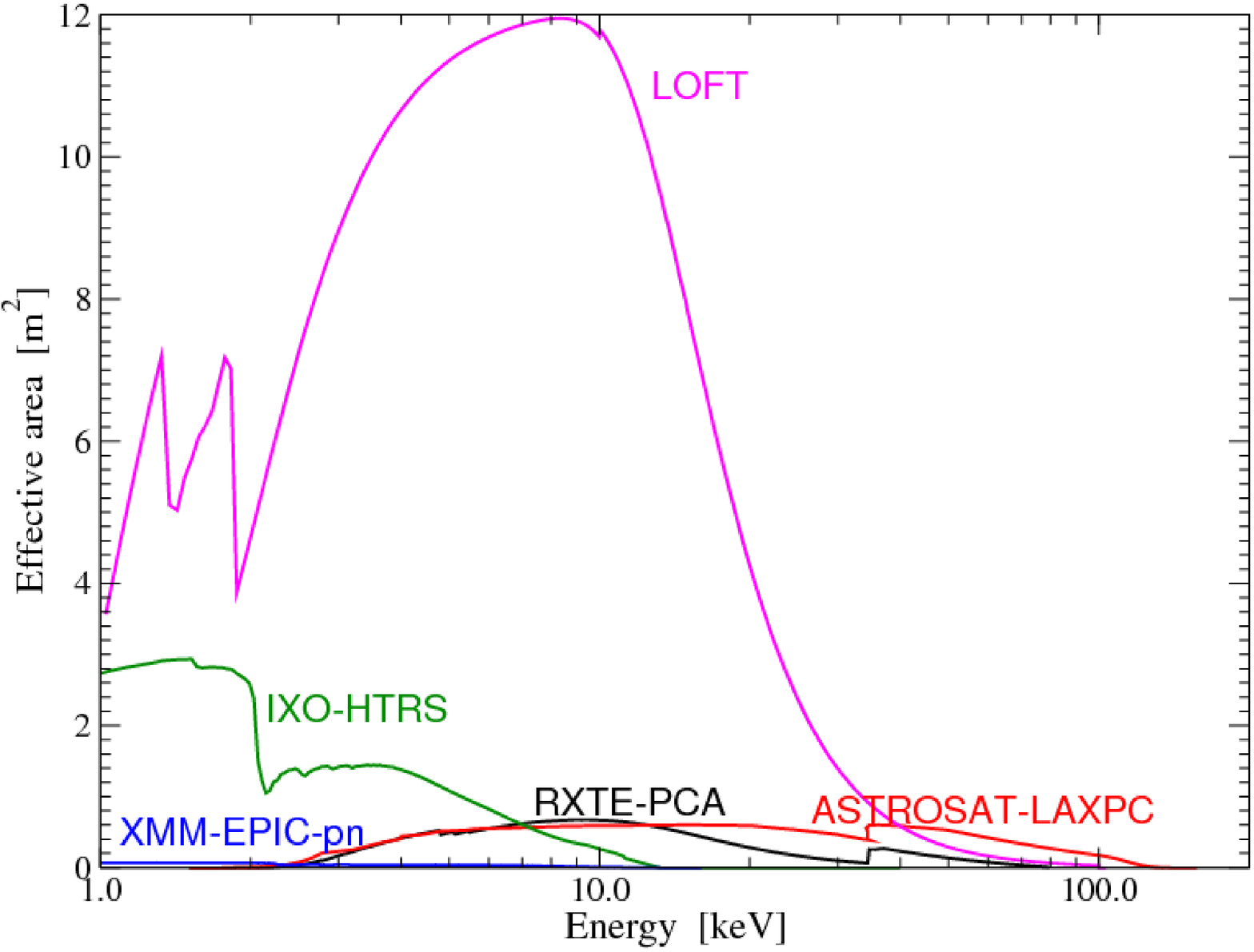} 
 \caption{Left: Conceptual scheme of the {\em LOFT} satellite: Large Area Detector (green), Wide Field Monitor (yellow), optical bench (red), structural tower (purple), bus (gold) and solar array (blue). Right: LAD effective area (vs. energy) plotted in linear  scale, as compared to that of other satellites for X-ray astronomy (from Feroci et al.\ 2011a)}
   \label{fig1}
\end{center}
\end{figure}

\section{Payload}

{\em LOFT} will be launched on a Vega rocket in a $\sim$ 600 km equatorial orbit. It will carry two instruments: a Large Area Detector (LAD), operating in the 2--50 keV range with energy resolution $<$260 eV (@6 keV) and a Wide Field Monitor (WFM). The LAD consists of 6 panels deployable in space (Fig. 1, left) which provide a total effective area of $\sim$ 10 m$^2$ (@10  keV), improving by a factor of $\sim$20  over its predecessors (Fig.1, right). The ground-breaking characteristic of the LAD is a mass per unit area of $\sim$10 kg/m$^2$, a factor of 10 lower than the {\em RXTE}/PCA, enabling a $\sim$10 m$^2$ area payload at reasonable weight. The ingredients for a sensitive but light experiment are the large-area Silicon Drift Detectors, and a collimator based on lead-glass micro-channel plates. An unprecedently large throughput ($\sim 3\times10^{5}$ cts/s from the Crab) will be achieved, while making pile-up and dead-time secondary issues.  The WFM is a coded-mask telescope mounted at the top of the structural tower at the centre of the LAD deployable array. The WFM  will operate in the energy range 2--50 keV and with a field of view of 3 steradians, corresponding to $\sim$1/4 of the whole sky. The WFM  angular resolution (5 arcmin) will enable it to locate sources with a 1 arcmin accuracy, with a $5\sigma$ sensitivity of 2 mCrab (50 ks).

\begin{table}
  \begin{center}
  \caption{Overview of the {\em LOFT} instrument performances.}
  \label{tab1}
 {\scriptsize
  \begin{tabular}{lll}\hline 
{\bf Item} & {\bf Requirement} & {\bf Goal}  \\ \hline
\multicolumn{3}{c}{Large Area Detector (LAD)} \\ \hline
Energy Range & 2--50 keV & 1--50 keV\\
Effective Area (2--10 keV) & 10 m$^2$ @ 8 keV& 12 m$^2$ @ 8 keV\\
Energy Resolution (@ 6 keV) & 260 ev @ 6 keV & 200 ev @ 6 keV \\
Field of View (FWHM) & $< 1^{\circ}$; transparency $<$1\% @ 20 keV& 30 arcmin \\
Time Resolution& 10$\mu$s & 7$\mu$s \\
Dead Time& $<$1\% (@1 Crab$^1$) & $<$0.5\% (@1 Crab)\\
Background& $<$ 10 mCrab& $<$ 5 mCrab\\ 
Maximum source flux (steady,peak) &$>$500mCrab; $>$15 Crab & $>$500mCrab; $>$30 Crab\\ \hline
 \multicolumn{3}{c}{Wide Field Monitor (WFM)} \\ \hline
Energy Range 2--50 keV & 1--50 keV\\
Energy Resolution (FWHM) &500 eV & 300 eV\\
Field of View & 50\% of the accessible LAD sky coverage& Same, with improved sensitivity\\
Angular Resolution & 5 arcmin  & 3 arcmin\\
Point Source Localisation & 1 arcmin & 0.5 arcmin\\
Sensitivity ($5\sigma$, 50 ks) & 5 mCrab& 2mCrab\\ \hline
  \end{tabular}
  }
 \end{center}
 \scriptsize{
{\it Notes:} $^1$Flux values are in units of the Crab pulsar flux in the energy range of interest.\\}
\end{table}

\section{The LOFT Science Driver: study of matter under extreme conditions}

Science drivers for {\em LOFT} are the study of  the  neutron star structure and the equation of state (EOS) of ultra-dense matter (mass,  radius and crustal properties of neutron stars),  the motion of matter under strong gravity conditions and the mass and spin of the black holes via the study of quasi periodic oscillations (QPOs) in the time domain, relativistic precession, Fe line reverberation studies in AGNs, the measure of small amplitude periodicities in X-ray transients,  millisecond pulsars, etc., discovery of new X-ray transients, early trigger of jets over many astronomical scales X-ray flashes, and many others. 
The LAD $\sim$10 m$^2$ effective area  in the 2-50 keV energy range will allow timing measurements of unprecedented sensitivity, leading for instance  the measure the mass and radius of neutron stars with ~5\% accuracy, or to reveal blobs orbiting close to the marginally stable orbit in active galactic nuclei.The LAD energy resolution will also allow the simultaneous exploitation of spectral diagnostics, in particular from the relativistically broadened 6-7 keV Fe-K lines. The  WFM will monitor a large fraction of the sky and constitute an important resource in its own right. The WFM will discover and localise X-ray transients and impulsive events and monitor spectral state changes with unprecedented sensitivity. It will then trigger follow-up pointed observations with the LAD as well as with other multi-wavelength facilities.

\end{document}